\documentclass{amsproc}
\usepackage{oldgerm}
\usepackage{latexsym}
\usepackage{amsmath}
\usepackage{amssymb}
\usepackage{amsgen}
\usepackage{amscd} 
\usepackage{color}
\usepackage{epsfig}

\theoremstyle{definition}

\theoremstyle{remark}

\theoremstyle{remark}

\title[On the use of photonic $N00N$ states]  
{On the use of photonic $N00N$ states for\\ 
practical quantum interferometry$^\dag$}
\author[Gerald Gilbert, Michael Hamrick and
Yaakov S. Weinstein]{Gerald Gilbert, Michael Hamrick and
Yaakov S. Weinstein \\
\small \it Quantum Information Science Group$^\ast$$^\ddag$\\
{\sc  Mitre} \\
\small \it 260 Industrial Way West, Eatontown, NJ 07724 USA
\footnote{$^\dag$Submitted on 22 November 2006 to DARPA for public release approval.
Approved on 18 December 2006 for Public Release, Distribution Unlimited.
}
\footnote{$^\ast$Research supported by DARPA under MITRE Project 0706D070-QS.
}
\footnote{$^\ddag$E-mail address: \tt{\{ggilbert, mhamrick, weinstein\}@mitre.org}}
}

\makeindex 

\begin{document} 

\begin{abstract}

The performance of photonic $N00N$ states, propagating in an attenuating medium, is analyzed
with respect to phase estimation. It is shown that, for $N00N$ states propagating through a 
lossy medium, the Heisenberg limit is never achieved. It is also shown that, for a given value of $N$, a signal comprised of an attenuated separable state of $N$ photons will actually produce a {\em better} phase estimate than will a signal comprised of an equally attenuated $N00N$ state, unless the transmittance of the medium is very high. This is a consequence of the need to utilize measurement operators appropriate to the different signal states. The result is that, for most practical applications in realistic scenarios with attenuation, the resolution of $N00N$ state-based phase estimation not only does not achieve the Heisenberg Limit, but is actually {\em worse} than the Standard Quantum Limit. It is demonstrated that this performance deficit becomes more pronounced  as the number, $N$, of photons in the signal increases. 

\end{abstract}

\maketitle

\newpage

\phantom{aa}

\newpage

%\tableofcontents

\section{Introduction}

Research performed over the past several years has suggested that quantum interferometric sensors that make use of certain entangled states can achieve phase estimation results that are superior to those that can be achieved using separable states \cite{SM,D1,D2,Seth1}. Entanglement correlations can be exploited in such a way that the Standard Quantum Limit on the estimation of phase, the error for which is given by $1/\sqrt{N}$, can be replaced with the Heisenberg Limit, the error for which is given by $1/N$, where $N$ is the number of photons in the signal state. This improvement in phase estimate precision of a factor of $1/\sqrt{N}$ has led to suggestions of a wide variety of prospective applications. In \cite{Boto} a specific method was presented that makes use of $N00N$ states to achieve arbitrarily good precision in the estimation of phase, increasing with the value of $N$, which was suggested as the basis for interferometric photolithography, known as ``quantum lithography" \cite{QL1,QL2}. Other applications using the proposed method have since been suggested such as quantum clock-synchronization and geodesy \cite{Seth2,QCS}, quantum imaging \cite{QI1,QI2} and others \cite{SethRev}. These applications are enabled by 
$N00N$ state-based interferometric protocols that beat the Standard Quantum Limit, and achieve the Heisenberg Limit. However, these results were established for signals propagating in perfect devices, and through lossless media \cite{pre}. In this paper we consider in detail the effect of attenuation on the precision of phase estimation. We find dramatic degradation in performance that calls into question the suitability of $N00N$ states 
for practical quantum interferometry.

The outline of the paper is as follows. In Section 2 we review the established results appropriate to lossless propagation of $N00N$ states. In Section 3 we explicitly calculate the modifications of $N00N$ state performance that arise due to attenuation. In Section 4 we carry out a comparison of the performance of attenuated $N00N$ states with the performance of equally attenuated $N$-photon separable states. Section 5 contains a discussion of the implications of these results, including comments on the notion that quantum computational error correction techniques can ameliorate the degradation in phase estimation precision that attenuation produces.

\section{Phase Error for $N00N$ States in Absence of Attenuation}

The initial $N00N$ state is given by
\begin{eqnarray}
\label{noonstate}
|\psi_{N00N}\rangle 
  &=& \frac{1}{\sqrt 2} \left( |N0\rangle  + |0N\rangle \right) \nonumber\\
  &=& \frac{1}{\sqrt 2} 
     \left( \frac{a_1^{\dag N}}{\sqrt{N!}} + \frac{a_2^{\dag N}}{\sqrt{N!}} \right)
     |v\rangle~,
\end{eqnarray}
where $|v\rangle$ is the vacuum state and $a_i^\dag$, $i=1,2$, is the photon creation operator 
for arm $i$ of the interferometer that will be used to perform the phase estimation.

The phase shift in the second arm of the interferometer is characterized by its effect
on the field operator:
\begin{equation}
a_2 \mapsto e^{-i\phi} a_2 ~,
\end{equation} 
where $\phi$ is the phase shift we wish to estimate.  Applying the 
phase shift operation to the $N00N$ state, $|\psi_{N00N}\rangle$, we obtain the phase-shifted 
$N00N$ state, $|\psi^\prime_{N00N}\rangle$, as
\begin{eqnarray}
\label{noonstate2}
|\psi^\prime_{N00N}\rangle 
  &=& \frac{1}{\sqrt 2} 
     \left( \frac{a_1^{\dag N}}{\sqrt{N!}} + 
            e^{iN\phi} \frac{a_2^{\dag N}}{\sqrt{N!}} \right)
     |v\rangle \nonumber\\
  &=& \frac{1}{\sqrt 2} \left( |N0\rangle  + e^{iN\phi} |0N\rangle \right)~.
\end{eqnarray}
We estimate the phase $\phi$ by performing a measurement of a suitable observable 
on this state.  Following \cite{QRS} we use the observable given by 
\begin{eqnarray}
A_D &=& |N0\rangle \langle 0N| + |0N\rangle \langle N0| \nonumber\\
    &=& \frac{1}{N!} \left( 
       a_1^{\dag N} |v\rangle \langle v| a_2^N + 
       a_2^{\dag N} |v\rangle \langle v| a_1^N \right) ~.
\end{eqnarray}
With this choice of phase estimation observable we obtain the measurement noise 
\begin{eqnarray}
\Delta A_D &\equiv& \sqrt{\langle A_D^2 \rangle - \langle A_D \rangle^2} \nonumber\\
  &=& \vert\sin{N\phi}\vert
\end{eqnarray}
and phase responsivity
\begin{equation}
\frac{d\langle A_D \rangle}{d\phi} = 
           -N \sin{N\phi}~,
\end{equation}
from which the phase error is obtained as
\begin{eqnarray}\label{diff}
\delta\phi &\simeq& \frac{\Delta A_D}
         {\vert \frac{d\langle A_D \rangle}{d\phi} \vert} \\
&=& \frac{1}{N}~.
\end{eqnarray}
Thus, the phase measurement under lossless conditions achieves the Heisenberg limit $1/N$.

\section{Dramatic Increase in Phase Error for Attenuated $N00N$ States}

The above analysis was carried out under the assumption that the $N00N$ states propagate 
through perfectly lossless media. However, any practical realization of quantum sensing 
will necessarily involve losses. In particular, the interferometer will be imperfect, 
resulting in some degradation of the signal states. We now modify the above analysis to take 
into account the attenuation in the two arms of the interferometer.  
As before, we begin with the initial $N00N$ state given by \eqref{noonstate}. Since the $N00N$ states in this paper are presumed to be realized by photons, we make use of the model for photon attenuation given in \cite{loudon}, in terms of which the effect of the attenuation on the field operators is given by
\begin{equation}
a_i \mapsto e^{\left(i \eta_i\omega/c - K_i/2 \right) L_i} a_i + 
   i \sqrt{K_i} \int_0^{L_i} dz 
       e^{\left(i \eta_i\omega/c - K_i/2 \right) \left( L_i-z \right)} b(z)~,
\end{equation}
where $\eta_i$ is the index of refraction for arm $i$ of the interferometer, 
$K_i$ is the attenuation coefficient, and $L_i$ is the path length.  The field 
operator function $b(z)$ represents modes into which photons are scattered by 
attenuation processes \cite{fieldop}.  As before, we introduce the phase shift in the second 
arm of the interferometer by 
\begin{equation}
a_2 \mapsto e^{-i\phi} a_2 ~,
\end{equation} 
where $\phi$ is the phase shift we wish to estimate.  
Applying the phase shift and taking account of attenuation, the $N00N$ state becomes
\begin{eqnarray}
|\psi^\prime_{N00N}\rangle && = \frac{1}{\sqrt{2 N!}}
     \lbrack 
     e^{\left( -i\eta_1\omega/c - K_1/2 \right)NL_1} a_1^{\dag N} + \nonumber\\
     && \phantom{aa}
     e^{\left( -i\eta_2\omega/c - K_2/2 \right)NL_2} e^{iN\phi} a_2^{\dag N} +
     \cdots
     \rbrack |v\rangle ~,
\end{eqnarray}
where the ellipsis 
refers to states outside the ${|N0\rangle, |0N\rangle}$ basis which are lost to 
attenuation.  

As for the lossless case, the phase estimation observable is
\begin{eqnarray}
A_D &=& |0N\rangle \langle N0| + |N0\rangle \langle 0N| \nonumber\\
    &=& \frac{1}{N!} \left( 
       a_1^{\dag N} |v\rangle \langle v| a_2^N + 
       a_2^{\dag N} |v\rangle \langle v| a_1^N \right) ~,
\end{eqnarray}
for which we now obtain
\begin{eqnarray}
\Delta A_D &=& \lbrack \frac{1}{2} 
           \left( \alpha_1^N - 2\alpha_1^N \alpha_2^N + \alpha_2^N \right) \nonumber\\
           && \phantom{a} +
           \left( \alpha_1 \alpha_2 \right)^N \sin^2{N(\phi-\phi_0)} \rbrack^{1/2}~
\end{eqnarray}

\noindent and

\begin{eqnarray}
\frac{d\langle A_D \rangle}{d\phi} &=& 
           -N \left( \alpha_1 \alpha_2 \right)^{N/2} \sin{N(\phi-\phi_0)}~,
\end{eqnarray}
\noindent where we have introduced the transmittances
\begin{eqnarray}
\alpha_n &\equiv& e^{- K_n L_n}
\end{eqnarray}

\noindent and the dispersion shift

\begin{eqnarray}
\phi_0 &\equiv& \frac{\omega}{c} \left( \eta_2 L_2 - \eta_1 L_1 \right)~.
\end{eqnarray}

\noindent The transmittances, $\alpha_n$, $n=1,2$, for the two arms of the interferometer, are defined to be equal (in linear units) to unity when there is no signal loss, and equal to zero when there is complete loss of signal. 

The resulting phase error is~\cite{chen2}
\begin{equation}\label{dphi}
\delta\phi = \frac
   {\sqrt{\frac{1}{2} \left( \frac{1}{\alpha_1^N} - 2 + \frac{1}{\alpha_2^N} \right) +
      \sin^2{N(\phi-\phi_0)} }}
   {N \cdot \vert \sin{N(\phi-\phi_0)} \vert}~.  
\end{equation}
We see that this reduces to the Heisenberg limit, $1/N$, if there is no attenuation, that is, 
if $\alpha_1 = \alpha_2 = 1$. If either $\alpha_1\neq 1$ or $\alpha_2\neq 1$, the Heisenberg limit cannot be achieved. We observe that the derivation of eq.\eqref{dphi} makes use of the approximation to the derivative that appears in eq.\eqref{diff}. This approximation does not hold well in the regions where $\sin N(\phi-\phi_0)\approx 0$, for which the expression for $\delta\phi$ diverges. The exact, detailed behavior of the phase error function in these regions will be explored in a subsequent paper.

In practical applications of quantum sensing, one of the two interferometer arms (arm 1) will be in the device and/or laboratory, and the other arm (arm 2) will consist of the space through which the signal propagates in order to sense the phase object. Note that the phase shift applies physically to arm 2 of the 
interferometer, and this is the arm in which the photon encounters the phase 
object to be measured.  We call this the ``long" arm of the interferometer.  
Arm 1 of the interferometer does not directly contact the phase object, and 
consequently this arm can be fully enclosed in a controlled environment.  We refer to arm 
1 as the short arm.  In general, we anticipate that the transmittance of the 
short arm will be greater than the transmittance of the long arm, due to the fact 
that it is in a controlled environment. Furthermore, in cases where the phase object to be measured is a considerable distance from the measurement apparatus, the optical path can be much shorter for arm 1 and, for that reason as well, that arm 
is also likely to be in a more protected environment.

\begin{figure}
\begin{center}
\includegraphics[width = 8cm]{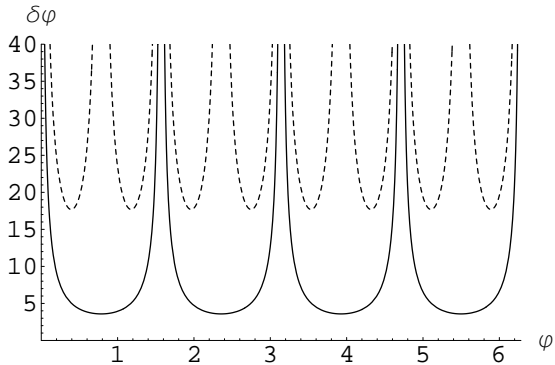}
\caption{\label{noon_atten}
Phase error for attenuated $N00N$ state.  Solid curve is for $N=2$, 
dashed curve is for $N=4$.  All values are in radians.}
\end{center}
\end{figure}

In Figure \ref{noon_atten}, we show the phase error for $N= 2,4$ with 
channel transmittances $\alpha_1 = 0.6$ and $\alpha_2 = 0.1$. The attenuation causes several new effects.  
The phase error is now a function of the phase and it is always 
much larger than the Heisenberg limit. The performance degrades rapidly with increasing photon number $N$ and with decreasing transmittance $\alpha_i$. The effect of even fairly modest levels of attenuation completely washes out any hoped-for improvement in resolution arising from the use of $N00N$ states that would have been obtained in the absence of attenuation.

\begin{figure}
\begin{center}
\includegraphics[width = 8cm]{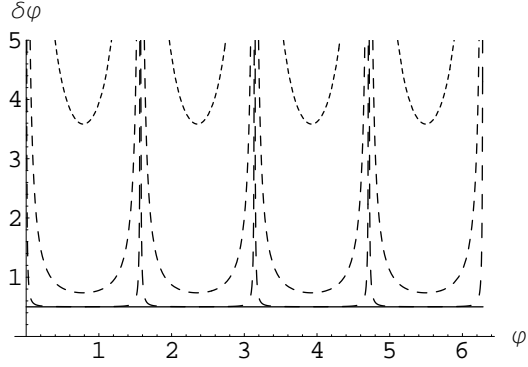}
\caption{\label{noon_limit_N=2}
Phase error in the limit of decreasing attenuation for $N00N$ states for $N=2$. 
  The uppermost (short dashed) curve is for $\alpha_1=0.6$, $\alpha_2=0.1$. The middle (medium dashed) curve is for $\alpha_1=0.8$, $\alpha_2=0.6$. 
The lower (long dashed) curve is for $\alpha_1=0.999999$, $\alpha_2=0.99$. The Heisenberg limit is denoted by the horizontal line at $\delta\phi=0.5$. Curve values in radians.}
\end{center}
\end{figure}

In Figure \ref{noon_limit_N=2}, we show the phase error for $N00N$ states with $N=2$ in the limit of decreasing attentuation for three sets of attenuation values for the arms of the interferometer. As the attenuation decreases, the phase error approaches, but never reaches, the Heisenberg limit, denoted by the horizontal line at $\delta\phi=0.5$. The uppermost (short dashed) curve is for $\alpha_1=0.6$, $\alpha_2=0.1$. The middle (medium dashed) curve is for $\alpha_1=0.8$, $\alpha_2=0.6$. 
The lower (long dashed) curve is for $\alpha_1=0.999999$, $\alpha_2=0.99$. The Heisenberg limit is achieved only for $\alpha_1=\alpha_2\equiv 1.$

\begin{figure}
\begin{center}
\includegraphics[width = 8cm]{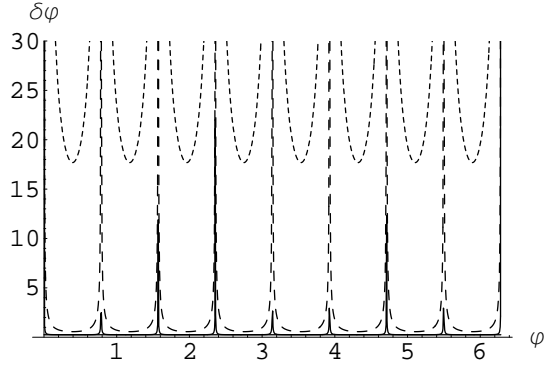}
\caption{\label{noon_limit_N=4}
Phase error in the limit of decreasing attentuation for $N00N$ states for $N=4$. 
This graph is to be compared with that in Figure \ref{noon_limit_N=2}. The attenuation values for the curves in this Figure for
the long and short arms of the interferometer are the same as for the corresponding curves
in Figure \ref{noon_limit_N=2}. The Heisenberg limit here appears at $\delta\phi=0.25$. Curve values in radians.}
\end{center}
\end{figure}

In Figure \ref{noon_limit_N=4}, we show the phase error for $N00N$ states with $N=4$ in the limit of decreasing attenuation for three sets of attenuation values for the arms of the interferometer. This graph for $N=4$ is to be compared with that in Figure \ref{noon_limit_N=2} which was calculated for $N=2$. The attenuation values for the curves in  Figure \ref{noon_limit_N=4} for the long and short arms of the interferometer are the same as for the corresponding curves in Figure \ref{noon_limit_N=2}. The Heisenberg limit here occurs at $\delta\phi=0.25$. This graph shows that the phase estimation performance for attenuated $N00N$ states becomes worse as $N$ increases, reflected by the increasing values of $\delta\phi$.

\section{Comparison of Separable vs. $N00N$ State $N$-Photon Results for 
Attenuative Channels}

In this section we compare the phase estimation performance of signals comprised of attenuated $N00N$ states with the phase estimation performance of signals comprised of attenuated separable states of $N$ photons. It is important to emphasize that carrying out an interferometric-based estimate of the phase requires choosing: (1) a signal state {\em and} (2) a measurement operator. We will refer to a choice of the pair \{{\em state}, {\em operator}\} as a phase estimation {\em method}. In the analysis of the previous section the \{{\em state}, {\em operator}\} pair consisted of 
$\{|\psi_{N00N}\rangle ,~
A_D \} \equiv \{\frac{1}{\sqrt 2} \left( |N0\rangle  + |0N\rangle \right) ,~
|N0\rangle \langle 0N| + |0N\rangle \langle N0|
\}$. We will henceforth refer to this choice as the ``$N00N$-state method" for phase estimation.

In this section we analyze the phase estimation performance of signals comprised of separable states of $N$ photons. The \{{\em state}, {\em operator}\} pair appropriate to this analysis is explicitly given below. We refer to this choice as the ``$N$ photon separable-state method." Below we calculate the phase errors for the $N$ photon separable-state method in the presence of attenuation. We use this result as a reference point for evaluating the performance of the $N00N$-state method in the presence of attenuation. If the $N00N$-state method performs poorly in comparison with the 
the $N$ photon separable-state method, there is 
no reason to expend the extra effort required to prepare and use the entangled 
$N00N$ state, as the $N$ photon separable-state method, based on states which 
are easily prepared and used, would provide superior phase estimates.  
In fact, the $N00N$-state method needs to achieve a significant 
improvement over the $N$ photon separable-state method to justify its use. We shall see that 
the $N00N$-state method achieves a substantial improvement only when the 
transmittance of the photon channels is very high.  In many cases, 
the required transmittances are likely to be difficult to achieve even under controlled
laboratory conditions.  

The initial separable state of $N$ photons is given by 
\begin{eqnarray}
\label{classicalstate}
|\psi_{N}\rangle 
  &=& \frac{1}{\sqrt {2^N}} \left( |10\rangle  + 
                |01\rangle \right)^{\otimes N}  \nonumber\\
  &=& \frac{1}{\sqrt {2^N}} \prod_{k=1}^N 
     \left( a_{k,1}^\dag + a_{k,2}^\dag \right)
     |v\rangle~,
\end{eqnarray}
where $|v\rangle$ is the vacuum state and $a_{k,i}^\dag$ 
creates photon $k$ 
in arm $i$ of the interferometer ~\cite{photonk}.  Attenuation and dispersion 
in the channel is given by
\begin{equation}
a_{k,i} \mapsto e^{\left(i \eta_i\omega/c - K_i/2 \right) L_i} a_{k,i} + 
   i \sqrt{K_i} \int_0^{L_i} dz 
       e^{\left(i \eta_i\omega/c - K_i/2 \right) \left( L_i-z \right)} b_k(z)~,
\end{equation}
and the phase shift in arm 2 is
\begin{equation}
a_{k,2} \mapsto e^{-i\phi} a_{k,2} ~.
\end{equation} 
The resulting state is 
\begin{eqnarray}
|\psi^\prime_N\rangle && = \frac{1}{\sqrt{2^N}} \prod_{k=1}^N 
     \lbrack 
     e^{\left( -i\eta_1\omega/c - K_1/2 \right)L_1} a_{k,1}^\dag + \nonumber\\
     && \phantom{aa}
     e^{\left( -i\eta_2\omega/c - K_2/2 \right)L_2} e^{i\phi} a_{k,2}^\dag +
     \cdots
     \rbrack |v\rangle ~,  
\end{eqnarray}
where the ellipsis, as in the analysis for $N00N$ states, refers to states lost to attenuation.  

The natural measurement observable for separable states of $N$ photons is
\begin{equation}
A_R \equiv \bigoplus_{k=1}^N A_R^{(k)}~,
\end{equation}
where
\begin{eqnarray}
A_R^{(k)} &=& |01{\rangle}_{kk}{\langle} 10| + |10\rangle_{kk}\langle 01| \nonumber\\
    &=& a_{k,1}^\dag  |v\rangle \langle v| a_{k,2} + 
       a_{k,2}^\dag |v\rangle \langle v| a_{k,1}~.  
\end{eqnarray}
\\

\noindent Thus, the $N$ photon separable-state method is defined by the pair
$\{|\psi_N\rangle ,~
A_R \} \equiv \{\frac{1}{\sqrt {2^N}} \left( |10\rangle + |01\rangle \right)^{\otimes N}
,~\bigoplus_{k=1}^N A_R^{(k)}
\}$.

\noindent The variance of the phase measurement is then
\begin{eqnarray}
\Delta A_R &=& \lbrack \frac{N}{2} 
           \left( \alpha_1 - 2\alpha_1 \alpha_2 + \alpha_2 \right) \nonumber\\
           && \phantom{a} +
           N\left( \alpha_1 \alpha_2 \right) \sin^2{(\phi-\phi_0)} \rbrack^{1/2} ~,
\end{eqnarray}
with the responsivity given by
\begin{equation}
\frac{d\langle A_D \rangle}{d\phi} = 
           -N \sqrt{ \alpha_1 \alpha_2} \sin{(\phi-\phi_0)}~.
\end{equation}
The resulting phase error is
\begin{equation}
\delta\phi = \frac
   {\sqrt{\frac{1}{2} \left( \frac{1}{\alpha_1} - 2 + \frac{1}{\alpha_2} \right) +
      \sin^2{(\phi-\phi_0)} }}
   {\sqrt{N} \cdot \vert \sin{(\phi-\phi_0)} \vert}~.  
\end{equation}
Note that this reduces to the Standard Quantum Limit, $1/\sqrt{N}$, if there is no 
attenuation, that is, 
if $\alpha_1 = \alpha_2 = 1$.  

The phase error is given by a periodic function of the phase, exhibiting minima for $\sin{(\phi-\phi_0)}=1$.  We will compare the 
performance of the $N00N$ state method with that of the $N$ photon separable-state method by comparing 
the values of the minima in $\delta\phi$.  The following graphs make this comparison 
for various values of $N$ and photon channel transmittance.  

\begin{figure}
\begin{center}
\includegraphics[width = 8cm]{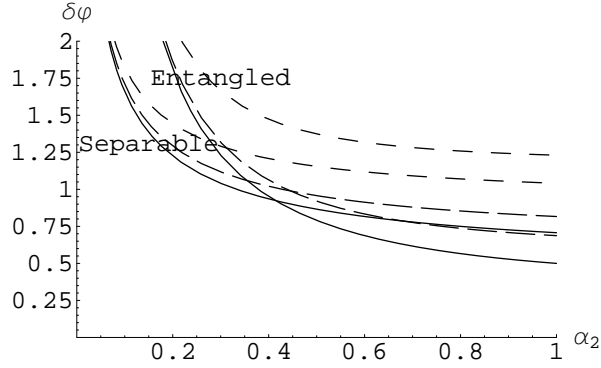}
\caption{\label{n2plot}
Comparison of minimum phase error for attenuated 
entangled $N00N$ states 
with minimum phase error for attenuated separable $N$-photon states. 
All curves are for $N=2$.  
The abscissa is the transmittance of the ``long" arm of the interferometer.   
For the solid curves, the transmittance in the ``short" arm is perfect: $\alpha_1=1.0$.  For the long-dashed curves the short arm transmittance $\alpha_1=0.6$.  For the short-dashed curves, $\alpha_1=0.3$.  
Phase errors are in radians.}
\end{center}
\end{figure}

Figure \ref{n2plot} compares the minimum phase error for 2 photon $N00N$ states 
to that for the $N$ photon separable-state method applied to separable 2-photon states for various 
levels of attenuation.  
It shows that the performance of the $N=2\phantom{a}N00N$ 
state degrades rapidly with attenuation as compared with the performance of 
the $N$ photon separable-state method.  
Even if we assume perfect transmittance for the short arm of the 
interferometer, the performance of the $N00N$ state is better than the 
separable state only for long arm 
transmittances $\alpha_2 \gtrsim 0.41$.  The long arm transmittance 
required for the $N00N$ state method to break even increases 
as the short arm transmittance decreases.  When the short arm 
transmittance $\alpha_1 \lesssim 
0.41$, the 
$N00N$ state performance is {\em always} worse than the separable state.  

\begin{figure}
\begin{center}
\includegraphics[width = 8cm]{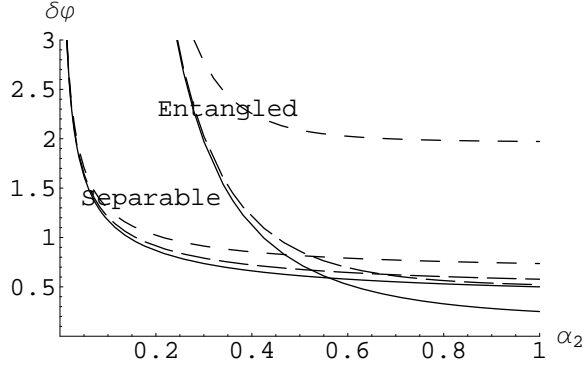}
\caption{\label{n4plot}
Comparison of minimum phase error for attenuated 
entangled $N00N$ states 
with minimum phase error for attenuated separable $N$-photon states. 
All curves are for $N=4$.  
The abscissa is the transmittance of the long arm of the interferometer.   
For the solid curves, the transmittance in the short arm is perfect: $\alpha_1=1.0$.  For the long-dashed curves the short arm transmittance $\alpha_1=0.6$.  For the short-dashed curves, $\alpha_1=0.3$.  
Phase errors are in radians.}
\end{center}
\end{figure}

Figure \ref{n4plot} compares the minimum phase error for 4 photon $N00N$ states 
to that for the $N$ photon separable-state method.  The interpretation of the 
results is analogous to that for the 2 photon case.  
The performance of the $N=4\phantom{a}N00N$ 
state degrades rapidly with attenuation as compared with the performance of 
the $N$ photon separable-state method, and the transmittances required for the $N00N$ state 
method to outperform the $N$ photon separable-state method are higher for $N=4$ than for $N=2$.   
If we assume perfect transmittance for the short arm of the 
interferometer, the performance of the $N00N$ state is better than the 
separable state for long arm 
transmittances $\alpha_2 \gtrsim 0.56$.  The long arm transmittance 
required for the $N00N$ state method to break even increases 
as the short arm transmittance decreases.  When the short arm 
transmittance $\alpha_1 \lesssim 
0.56$, the 
$N00N$ state performance is always worse than the separable state.

\begin{figure}
\begin{center}
\includegraphics[width = 8cm]{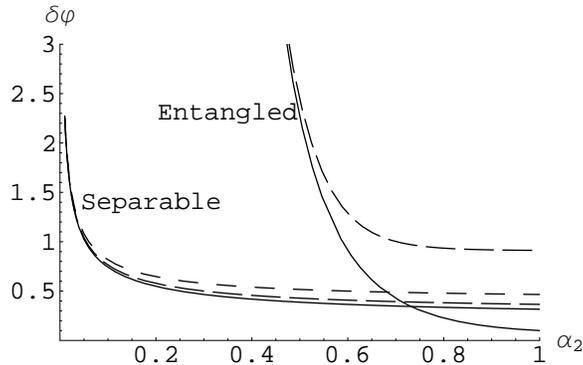}
\caption{\label{n10plot}
Comparison of minimum phase error for attenuated 
entangled $N00N$ states 
with minimum phase error for attenuated separable $N$-photon states. 
All curves are for $N=10$.  
The abscissa is the transmittance of the long arm of the interferometer.   
For the solid curves, the transmittance in the short arm is perfect: $\alpha_1=1.0$.  For the long-dashed curves the short arm transmittance $\alpha_1=0.6$.  For the short-dashed curves, $\alpha_1=0.3$.  
Phase errors are in radians.}
\end{center}
\end{figure}

Figure \ref{n10plot} compares the minimum phase error for 10 photon $N00N$ states 
to that for the $N$ photon separable-state method.  This is an important case for evaluating the practical 
feasibility of applying the $N00N$ state method outside of laboratory environments, 
since the improvement in performance is substantial when the transmittance is perfect, and 
it consequently appears to make sense to make the additional investment to prepare and use 
highly entangled states.   
As before, the figure shows that the performance of the $N=10\phantom{a}N00N$ 
state degrades rapidly with attenuation as compared with the performance of 
the $N$ photon separable-state method, and the transmittances required for the $N00N$ state 
method to outperform the $N$ photon separable-state method are higher for $N=10$ than for $N=4$ or $N=2$.  
If we assume perfect transmittance for the short arm of the 
interferometer, the performance of the $N00N$ state is better than the 
separable state for long arm 
transmittances $\alpha_2 \gtrsim 0.73$.  The long arm transmittance 
required for the $N00N$ state method to break even increases 
as the short arm transmittance decreases.  When the short arm 
transmittance $\alpha_1 \lesssim 
0.73$, the 
$N00N$ state performance is always worse than the separable state.  

\begin{figure}
\begin{center}
\includegraphics[width = 8cm]{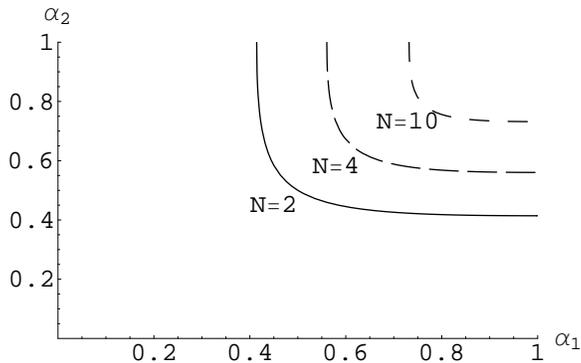}
\caption{\label{brkplot}
Transmittance curves for which the $N00N$ state method breaks even 
with the $N$ photon separable-state method in terms of the minimum phase error.  The $N00N$ state  performance 
is superior in the region above and to the right of the curves.}
\end{center}
\end{figure}

Figure \ref{brkplot} shows the values of transmittance in the short ($\alpha_1$) and long 
($\alpha_2$) arms 
for which the $N00N$ state method breaks even 
with the $N$ photon separable-state method in terms of the minimum phase error.  The $N00N$ state  performance 
is superior in the region above and to the right of the curves.  The figure shows that 
the practical 
applicability of the $N00N$ state method is limited to cases of very good transmittance.

\section{Conclusion}

In this paper we have analyzed the performance of photonic $N00N$ states propagating in an attenuating medium, such as would be present in realistic implementations of quantum interferometry. We find that even a modest amount of attenuation causes a dramatic degradation in phase estimation precision, calling into question the suitability of $N00N$ states for practical quantum interferometry. In fact the Heisenberg limit is never achieved by photonic $N00N$ states in the presence of any finite amount of attenuation. We have also shown that, unless the transmittance of the medium is unrealistically high, the phase estimation precision yielded by an attenuated $N00N$ state is actually {\em worse} than that obtained by using an equally attenuated separable $N$-photon state. Thus, for most practical applications in realistic scenarios, not only do $N00N$ states not achieve the Heisenberg Limit, but they actually fail to achieve the Standard Quantum Limit as well.

The intended purpose of applied quantum interferometry is to achieve significant improvement in the precision with which the phase can be estimated. The previously known results apply to an idealized scenario in which signals propagate through perfect devices and lossless media. These previous results indicate that increased precision in phase estimation can be achieved by increasing the value of $N$ in $N00N$ states. However, our results show that the $N00N$ state performance deficit caused by realistic attenuation becomes more pronounced as $N$ increases. In contrast to expectations based on the previous, idealized results, we find that the worst consequences of attenuation occur for precisely those values of $N$ for which the greatest phase resolution improvement was expected.

One might speculate that quantum error correction techniques taken from quantum computing theory could be applied to the problem of $N00N$ states degraded by attenuation in quantum interferometry. However, quite apart from questions involving the practical realization of such error correction techniques in a quantum interferometric context, there are reasons to suspect that error correction techniques will not be suitable for this purpose. In a quantum interferometric protocol, the phase object is interrogated to produce the phase estimate by establishing an interference pattern between the two arms of the interferometer in which the fiducial $N00N$ states propagate. The application of quantum error correction techniques would replace the fiducial $N00N$ states with {\em different}, encoded states, the resulting interference of which will in general produce a different phase estimate to the one associated to the fiducial signal. Moreover, in contrast to quantum interferometry, it should be noted that when applying quantum error correction to the construction of a fault-tolerant quantum computer, one is free to design encoded versions of the quantum gates so that they operate properly on encoded states.  This can not be done in quantum interferometry, since one is not free to ``re-engineer" the phase object that is being measured. 

A more fruitful approach to protecting against the consequences of attenuation in quantum interferometry might come from exploring the replacement of $N00N$ states with other entangled states that may prove more robust to the effects of attenuation, while still yielding improved phase estimation performance~\cite{GHW}.

\section{Acknowledgements}

\noindent We thank Michael Zatman and Esko Jaska for discussions.

%%%%%%%%%%%%%%%%%%%%%%%%%%%%%%%%%%%%%%%%%%%%%%%%%%%%%%%%%%%%%%%%%%%%%%%%%%%%%%%%%%%%%%
%%%%%%%%%%%%%%%%%%%%%%%%%%%%%%%%%%%%%%%%%%%%%%%%%%%%%%%%%%%%%%%%%%%%%%%%%%%%%%%%%%%%%%
\newpage
%-------------------------------------------------------------------------


\begin{thebibliography}{99}
\bibitem{pre}
The detailed analyses of $N00N$ state-based quantum interferometry of which we are aware have not explicitly calculated the degradation of phase estimate precision caused by attenuation. The analysis of \cite{chizhov} examined the propagation of entangled light pulses through absorbing and dispersing channels. Their analysis did not consider $N00N$ states, but analyzed, mainly for continuous-variable systems, the effect of dispersion and absorption on certain fidelity and correlation indices. They did not treat the question of phase estimate error in quantum interferometry.

The analysis of \cite{Chen} studied phase estimation by determining the relation between the precision achieved by attenuated $N00N$ states and the loss of entanglement associated to attenuation and phase damping. They did not directly compare the degradation in performance caused by attenuation with the Heisenberg and Standard Quantum limits, nor did they compare the performance of attenuated $N00N$ states with that of attenuated $N$-photon separable states. They also did not take account of dispersion loss in their analysis.
\bibitem{fieldop}
The field operator function, $b(z)$, has different physical dimensionality than do the 
field operators $a_i$, with the former related to the latter by a factor of a square root of length.
\bibitem{chen2}
Ref. \cite{Chen} also obtain an expression for the phase error using a different method. 
Their result and ours are in agreement for situations in which both apply.
\bibitem{photonk}
The photons in the separable state are injected into the interferometer at different times, 
labeled by $t_k$, where $k=1,\ldots ,N$, with measurements independently applied to each photon. 
The index $k$ is used to properly count these separate injection events.
\end{thebibliography}
\end{document}